\documentclass[twocolumn,floatfix,superscriptaddress,a4paper,showpacs,showkeys,nofootinbib,reprint,prc,noeprint]{revtex4-1}
\usepackage{epsfig}
\usepackage{latexsym}
\usepackage[colorlinks=true,linktocpage=true,linkcolor=blue,citecolor=blue,allcolors=blue]{hyperref}
\usepackage{url}
\usepackage[utf8]{inputenc}
\usepackage{enumerate}
\usepackage{color}
\usepackage{xcolor}
\usepackage{graphicx}
\usepackage{xfrac}

\usepackage{amsmath}
\usepackage{amssymb}

\usepackage[english]{babel}
\usepackage{hyperref}
\usepackage{url}
\usepackage{needspace}
\usepackage{float}
\hypersetup{
   colorlinks=true, 
  linktoc=all,     
  linkcolor=blue,  
}

\begin{document}

 \title{Hypernucleus production in p+Au reactions at the FAIR facility}

\author{Nitikorn Jaingarm}
\affiliation{Department of Physics, Naresuan University, Phitsanulok 65000, Thailand}

\author{Pornrad Srisawad}
\email{pornrads@nu.ac.th}
\affiliation{Department of Physics, Naresuan University, Phitsanulok 65000, Thailand}

\author{Christoph Herold}
\affiliation{School of Physics, and Center of Excellence in High Energy Physics and Astrophysics,
Suranaree University of Technology, Nakhon Ratchasima, 30000, Thailand}

\author{Ayut Limphirat}
\affiliation{School of Physics, and Center of Excellence in High Energy Physics and Astrophysics,
Suranaree University of Technology, Nakhon Ratchasima, 30000, Thailand}

\author{Jan Steinheimer}
\affiliation{GSI Helmholtzzentrum f\"ur Schwerionenforschung GmbH, Planckstr. 1, D-64291 Darmstadt, Germany}
\affiliation{Frankfurt Institute for Advanced Studies, Ruth-Moufang-Str. 1,  60438 Frankfurt am Main, Germany}

\author{Marcus Bleicher}
\affiliation{Institut f\"{u}r Theoretische Physik, Goethe-Universit\"{a}t Frankfurt, Max-von-Laue-Str. 1, D-60438 Frankfurt am Main, Germany}
\affiliation{Helmholtz Research Academy Hesse for FAIR (HFHF), GSI Helmholtzzentrum f\"ur Schwerionenforschung GmbH, Campus Frankfurt, Max-von-Laue-Str. 12, 60438 Frankfurt am Main, Germany}

\date{\today}

\begin{abstract}
We explore the production of hypernuclei in p+Au reactions using the UrQMD model accompanied by a standard phase space coalescence model. We focus on the proton beam energy range of $E_{\rm lab}= 5 - 30$ GeV as this energy range will be investigated by the CBM-experiment at the upcoming FAIR facility. Starting from proton, $\Lambda$, $\Sigma$, $\Xi$ and $\Omega$ production, we predict the yields, rapidity and transverse momentum distributions of $^{3}_{\Lambda}H$, $^{4}_{\Lambda}H$, $\Xi$N and $\Xi$NN hypernuclei. We conclude that the production rates of novel multi-strange hypernuclei are well within the reach of the CBM-experiment.
\end{abstract}

\maketitle
\section{Introduction}
Hypernuclei, i.e. atomic nuclei with additional (multi)strange baryons, are currently in the center of attention of high-energy heavy-ion studies. This novel type of nuclei can only be produced in a high-energy accelerators, where the initial collision energy allows to produce hyperons ($\Lambda, \Sigma, \Xi$ and possibly $\Omega$ baryons). Especially interesting are novel states of hypernuclei, with more than one hyperon, as they may allow to explore the hyperon-hyperon interaction \cite{Schaffner:1993qj,Schaffner-Bielich:1999zoi}. Generally, two competing effects govern the production rate of hypernuclei \cite{Akkelin:2001wv,Gaebel:2020wid}: I) The production rate of hyperons increases with beam energy, increasing the yields of hypernuclei, while on the other hand II) higher collision energies increase the freeze-out volume (lowering the baryon density) decreasing the probability of cluster production. It is well known that for nucleus-nucleus collisions the strange to light quark ratio has a maximum around 30 GeV beam energy \cite{Braun-Munzinger:2001alt}, indicating that one would also expect the maximal hypernucleus production in this energy range. For proton induced reactions this argument is altered, because the remaining cold nucleus may serve as an absorber to the produced hyperons without substantial expansion and may thus allow to explore large hypernuclei at lower beam energies.

Several experiments (ALICE \cite{ALICE:2024djx}, J-PARC \cite{He:2025vgg}, and STAR \cite{Zhou:2025efp}) have reported on the production of hypernuclei in high-energy collision. Especially, the new measurement at STAR \cite{Zhou:2025efp}, which has reported hypernuclei with $A=3-5$, and the pioneering experiments on double $\Lambda$ hypernuclei at J-PARC \cite{He:2025vgg} show interesting novel results. A new facility that will become available for these studies in the near future will be the Facility for Antiproton and Ion Research (FAIR), which will start operation in Darmstadt, Germany in 2028. A variety of experiments and initiatives are planned at FAIR to investigate hypernuclei: E.g. QCD@FAIR \cite{Messchendorp:2025men}, R3B \cite{Sun:2017unn,Ji:2025jzw}, and CBM \cite{CBM:2016kpk,Glaessel:2026rnq}. These complementary programs will study the structure of QCD matter with various experimental set-ups ranging from hadron-hadron reactions to nucleus-nucleus collisions. The expected discovery reach in the hypernucleus sector of these experiments can be estimated from the primary beam intensity of $10^{11}$/s for protons up to $E_{\rm kin}$=30 GeV and $10^{9}$/s for gold ions up to $E_{\rm kin}$=11$A$ GeV \cite{Selyuzhenkov:2020djo}, which will allow to produce approx. $10^{-1}-10^{-2}$ single strange hypernuclei and $10^{-4}$ double strange hypernuclei per Au+Au event \cite{Botvina:2014lga,Buyukcizmeci:2023azb,Buyukcizmeci:2024gpb}. Especially the QCD@FAIR program will pursue p+A collisions and is ideally suited for hypermatter studies due to three reasons: I) The initial beam energy is substantially higher than in A+A reactions at FAIR allowing for substantial production of hyperons, II) the massive target nucleus can absorb the produced hyperons without too much excitation yielding large fragments and III) the initial luminosity of the proton beam is higher, resulting in excellent statistics.

On the theoretical side, the formation of hypernuclei and the hyperon-nucleon (hyperon-hyperon) interaction \cite{Nagels:1976xq,Friedman:2022bpw,Yong:2025not,Jinno:2025mos} are relevant for the equation of state of neutron star matter \cite{Glendenning:1986wb,Maslov:2015wba,Jinno:2023xjr,Kochankovski:2025lqc} and the nuclear structure \cite{Knoll:2025vzp,Haidenbauer:2025zrr}. Hypernucleus production was investigated before in nucleus+nucleus reactions (see e.g. \cite{Botvina:2019lgn,Reichert:2022mek,Balassa:2023riq,Bratkovskaya:2025oys,Liu:2025kpp,Sun:2025oib}) using transport models supplemented by a coalescence mechanism. For multi-strange hypernuclei theoretical transport studies are more scarce, and we refer e.g. to  \cite{Schaffner-Bielich:1999zoi,Buyukcizmeci:2024gpb} where  hypernuclei with two $\Lambda$ hyperons and even more exotic multi-strange states were explored. Of special interest for the current study is the recent analysis of hypernucleus production in pion induced reaction at GSI \cite{Kittiratpattana:2023atz}. There, it was demonstrated that the residual nucleus acts as an effective decelerator and absorber of the produced hyperons and leads to copious production of large hypernuclei. This motivates our present study of p+A reactions.  

\begin{figure*}[t!] 
        \includegraphics[width=1.0\textwidth]{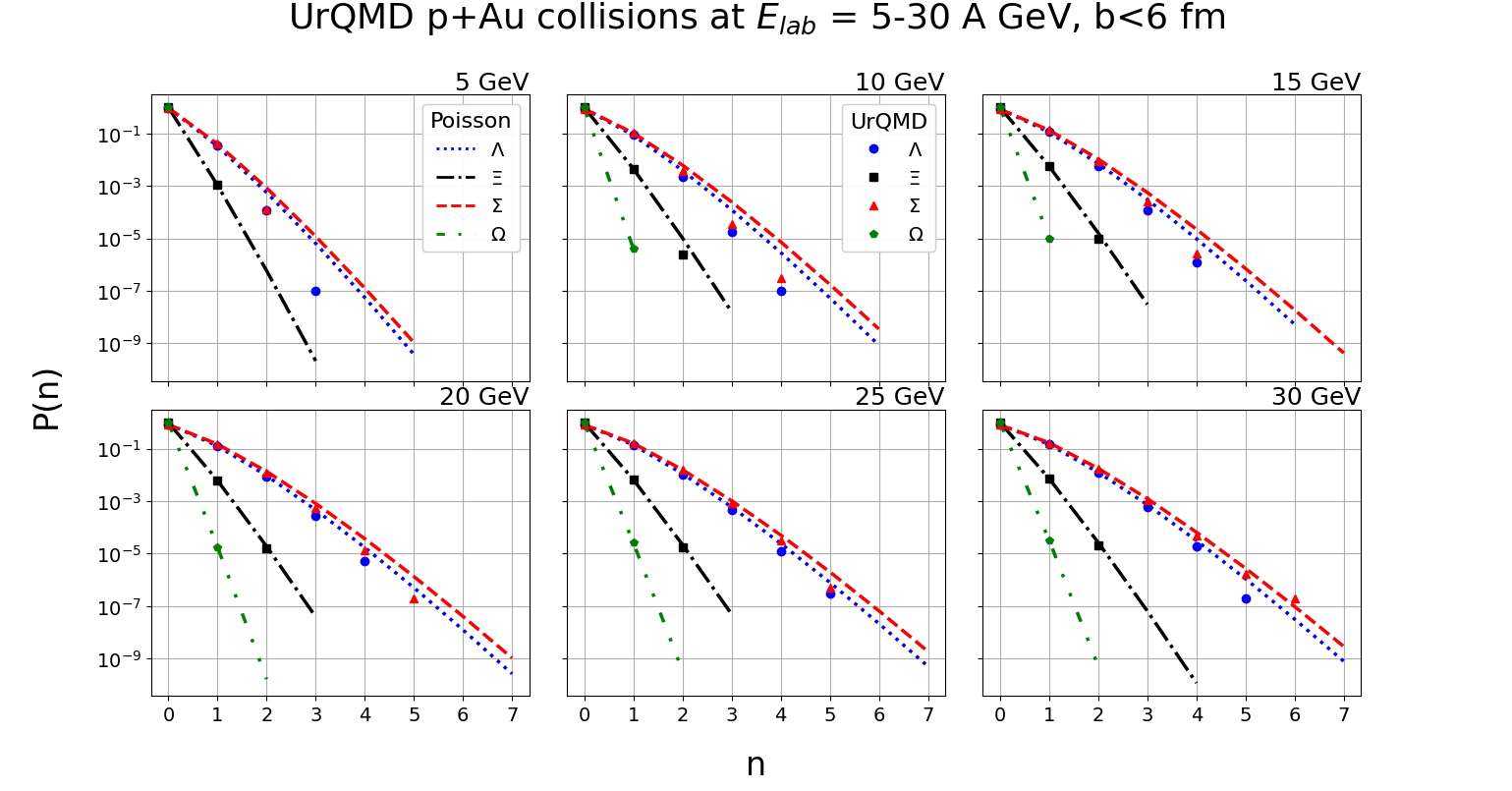}  
\caption{UrQMD calculation for min.bias p+A reaction in the energy range $E_{lab}$= 5-30 $A$GeV. Shown is the probability $P(n)$ to produce $n$ hyperons ($\Lambda$, $\Xi$, $\Sigma$, and $\Omega$) in a single event. The upper panel shows results for the beam energies $5$, $10$ and $15$ GeV, while the lower panel shows the energies $20$, $25$ and $30$ GeV. The symbols show the UrQMD results, the lines represent fits with a Poisson distribution to the UrQMD results.}
\label{fig:multiplicity_results}
\end{figure*}

\section{UrQMD model and coalescence}
For our studies we employ the Ultra-relativistic Quantum Molecular Dynamics (UrQMD) model \cite{Bass:1998ca,Bleicher:1999xi,Bleicher:2022kcu}. UrQMD is a microscopic transport model to simulate hadron-hadron, hadron-nucleus, and nucleus-nucleus collisions from the GSI-SIS energy range up to the top RHIC energies. It is designed as a multi-purpose tool that allows to study particle production and particle spectra and further serves as an input for multifragmentation models and hadron coalescence. The model is based on hadron degrees of freedom supplemented by strings for higher energies. The propagation (in case nuclear potentials are employed) follows the relativistic Hamilton equations of motion, without potentials, the hadrons move on straight line trajectories. Finally, UrQMD provides the time dependent phase space distributions of all included hadron and light cluster species.

To calculate (hyper-)clusters, a phase space coalescence mechanism has been included in UrQMD-4.0 (for details see \cite{Sombun:2018yqh, Reichert:2022mek}) which has been benchmarked and tested for a broad range of reactions in the past. The main features are: I) Coalescence is applied after the individual baryons kinetic freeze-out (i.e. after the last scattering/decay of the considered baryons). This is a good approximation to dynamical cluster formation \cite{Kireyeu:2023spj} as cluster that are formed earlier will be destroy from interaction and intense rescattering \cite{Oliinychenko:2018ugs}. II) If the baryons are sufficiently near in space and momentum space, they have a certain probability to form a cluster. The coalescence parameters are independent of energy and collision system and have been fixed in previous studies.

\section{Reach in hyperon multiplicity}
We start our investigation of min.bias\footnote{We define min.bias reactions as those with impact parameter $b<6$fm and remove events without interactions.} p+A reaction in the energy range $E_{lab}$= 5-30 $A$GeV by exploring the probability to produce multiple hyperons in the same event with UrQMD. This is of course a prerequisite for the production of multi-hyperon clusters. Fig. \ref{fig:multiplicity_results} shows the probability to produce $n$ hyperons ($\Lambda$, $\Sigma$, $\Xi$, and $\Omega$)\footnote{Here and in the rest of the paper, $\Sigma=\Sigma^-+\Sigma^0+\Sigma^+$, and $\Xi = \Xi^0 + \Xi^-$. $\Lambda$s do not include contributions from the $\Sigma^0$ decay.} in a single event in the range of beam energies discussed above. The upper panel shows results for the beam energies $5$, $10$ and $15$ GeV, while the lower panel shows the energies $20$, $25$ and $30$ GeV. The symbols show the UrQMD results, the lines represent fits with a Poisson distribution to estimate the reach in $n$.
 
We observe that there is a substantial probability to produce two or more hyperons in the same event. Especially in the case of $\Lambda$ and $\Sigma$ hyperons, events with up to 6 hyperons in a single event are possible (with the experimentally accessible probability). In the case of $\Xi$ hyperons, a realistic maximal reach up to three cascades can be expected. However, the $\Omega$ hyperon is produced only at the highest beam energy and producing two of them is so significantly suppressed that it seems unfeasible to search for double-$\Omega$ events. 

This calculation already shows that in this beam energy range there is a possibility to create hypernuclear clusters with more than one hyperon, or even combinations of different hyperons in one single nucleus. To estimate the actual production probability we will need to take into account the probability that these produced hyperons end up close in phase space to the other nucleons to form a hypernucleus.
 
\section{Rapidity and transverse momentum distributions of the nucleons and hyperons}
\begin{figure}[t!] 
        \includegraphics[width=0.5\textwidth]{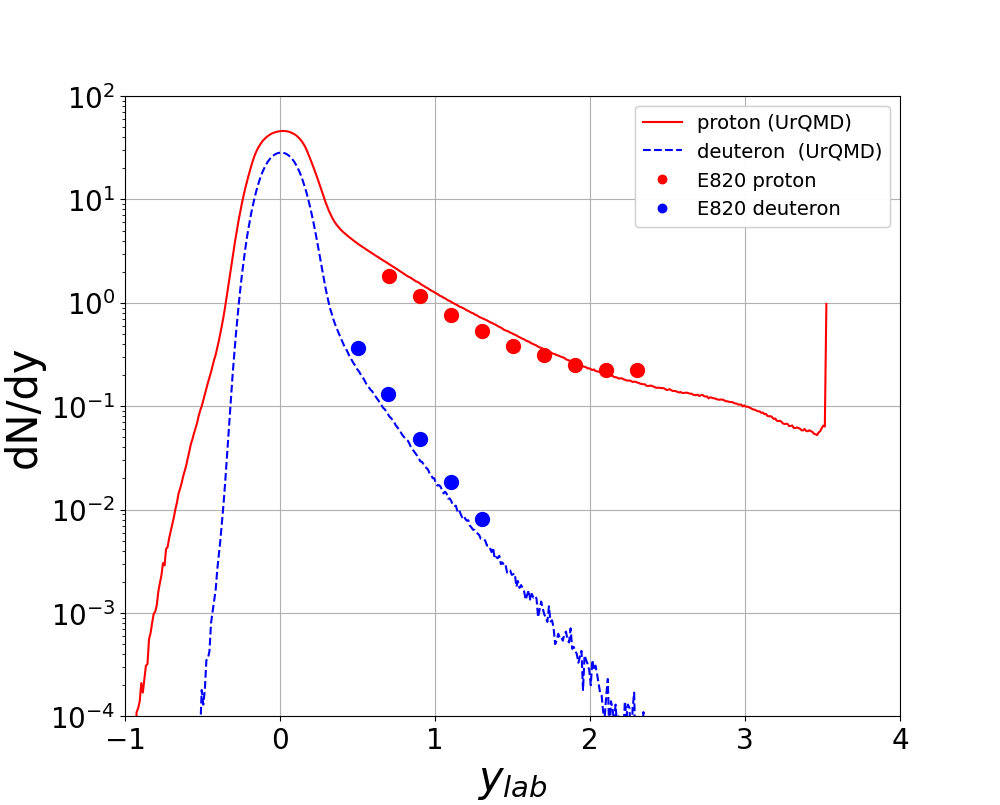}  
\caption{Rapidity density of protons (red) and deuterons (blue) in min.bias p+A reactions at $E_{\rm lab}=14.6$ GeV. Data from \cite{E-802:1991unu} are shown as symbols, UrQMD calculations are depicted by lines.}
\label{fig:pAu_data}
\end{figure}

As a first step, it is necessary to see if UrQMD produces also a correct n-particle phase space correlation of baryons in this asymmetric set-up. This can be tested (at least for the 2-particle correlation) by studying the production of deuterons on p+A reaction in this energy regime. In Fig. \ref{fig:pAu_data} we show the rapidity density of protons and deuterons in min.bias p+A reactions at $E_{\rm lab}=14.6$ GeV. Data from E802 \cite{E-802:1991unu} is shown as symbols. UrQMD calculations are depicted by lines. We observes that the model calculations provides a good description of the proton distribution (red symbols, red line) as well as the 2-particle correlations of the protons and neutrons leading to the production of deuterons (blue symbols, blue line). For the comparison of the 3-particle correlation (triton and ${}^3$He) to p+A data we refer to \cite{Hillmann:2021zgj}, although the data there is restricted to $p_T\approx 0$. In summary, we can assume that UrQMD provides a reasonable phase space correlation among the baryons to allow for quantitative predictions of hypernucleus production. 

\begin{figure*}[t!] 
    \centering
        \includegraphics[width=1.0\textwidth]{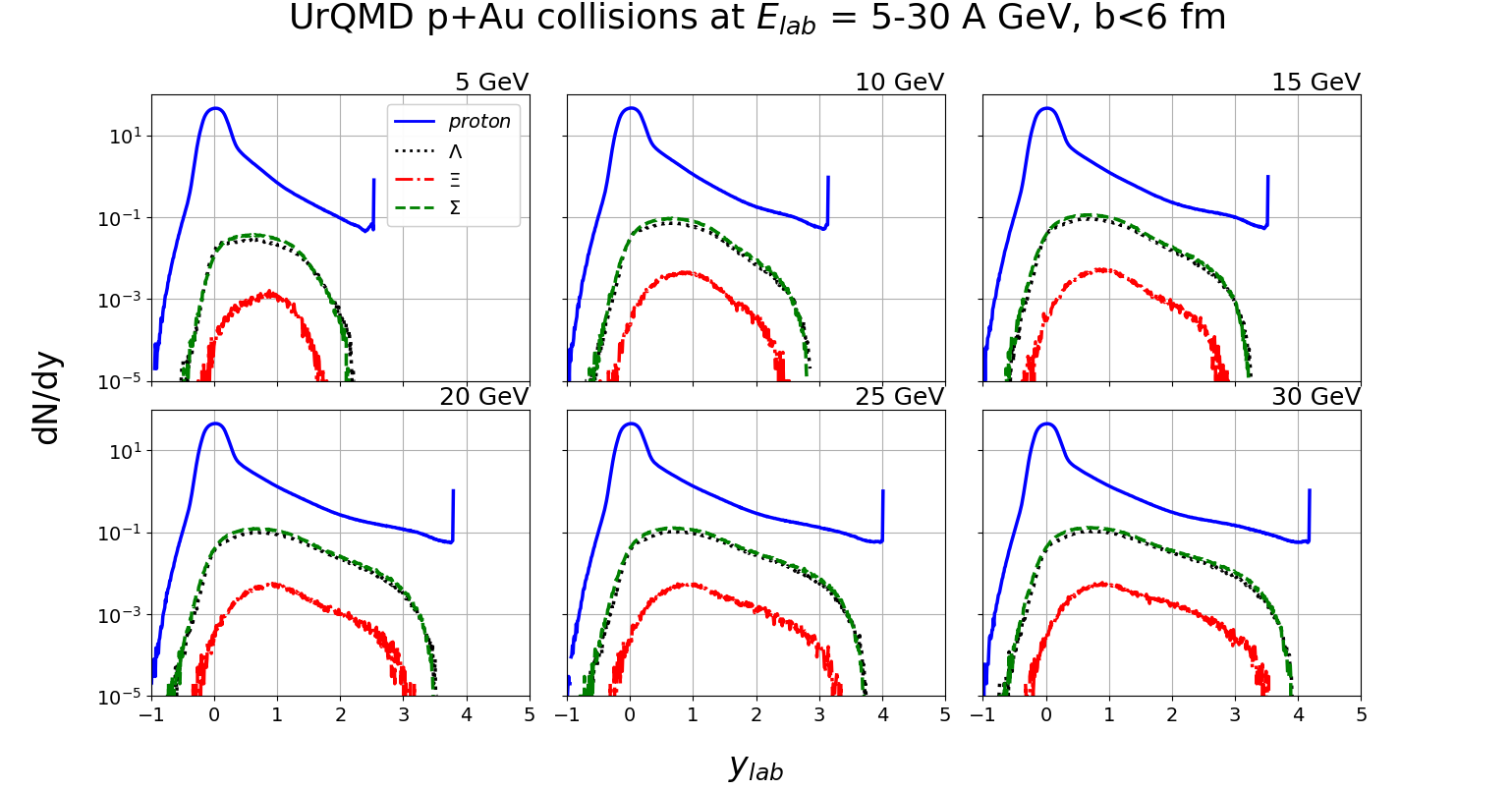}
    \caption{UrQMD calculation for the rapidity distribution of protons (blue) and hyperons ($\Lambda$ in black, $\Sigma$ in green, $\Xi$ in red) in the lab-frame in min.bias p+Au reactions at $E_{\rm lab}=$ 5-30 GeV.}
    \label{fig:dndy_hyperons}
\end{figure*}

To study the phase-space overlap of nucleons and hyperons we show in Fig. \ref{fig:dndy_hyperons} the rapidity distributions (in the laboratory frame, i.e. the target sits at $y=0$) of protons and various hyperons ($\Lambda$, $\Sigma$, $\Xi$,  and $\Omega$) for min.bias p+Au reactions at different beam energies of $E_{lab}$= 5-30 GeV. A cut in transverse momentum, similar to a typical experimental acceptance of $p_T>$0.3 GeV is applied. 

The peak of the distribution of protons is, as expected, around $y=0$ as this is where the target nucleus is located. Fewer protons get ejected in the forward direction from the projectile proton. The peak in rapidity for the hyperons is shifted forward by approx. half a unit of rapidity as they follow the motion of the center-of-mass system of the reaction. This indicates that there is significant overlap of the rapidity distributions of protons and hyperon's.
Such an overlap is a prerequisite to allow the formation of a nuclear cluster after the collision.


\begin{figure*}[t!] 
    \centering
        \includegraphics[width=1.0\textwidth]{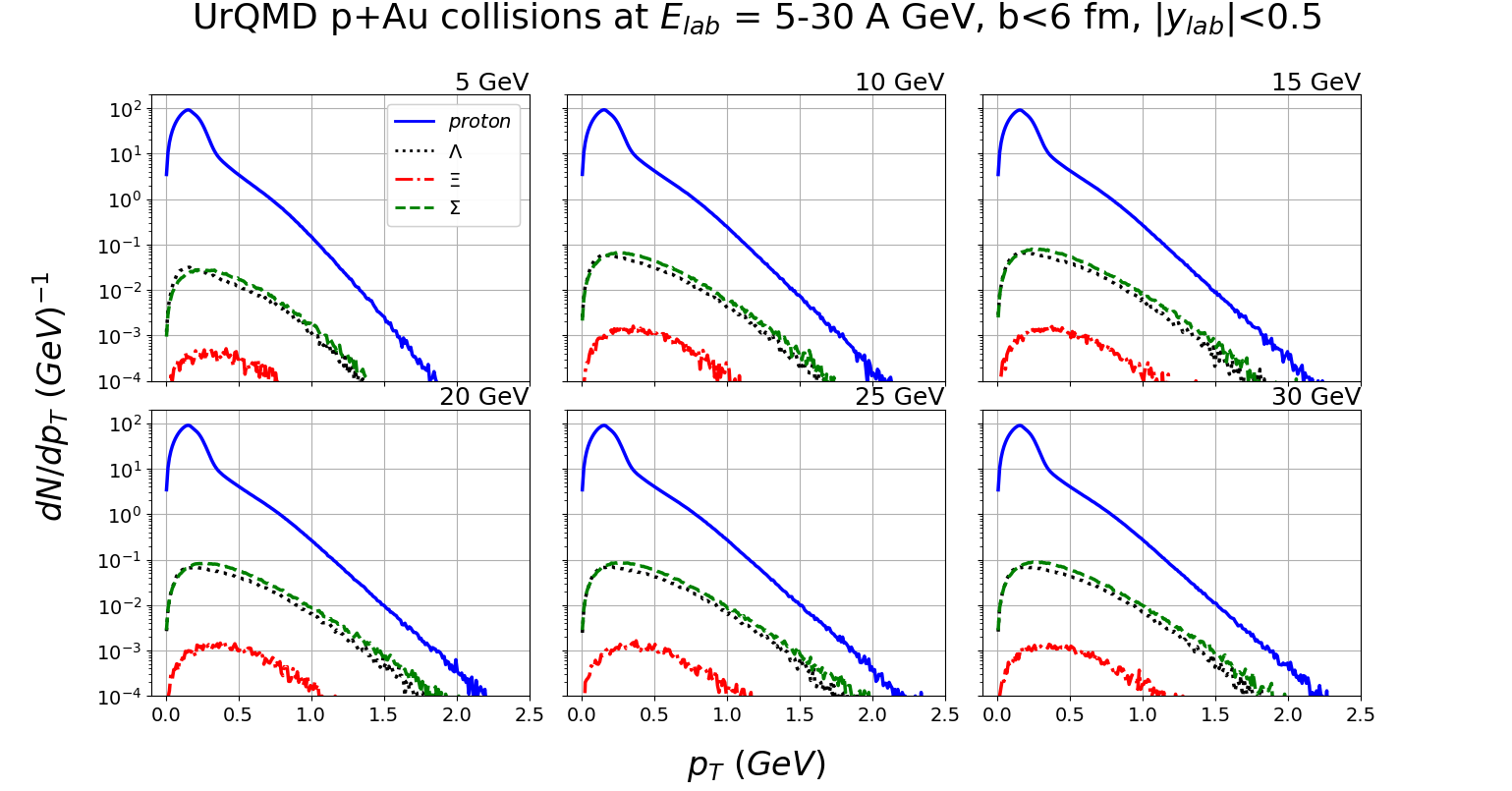}
    \caption{UrQMD calculation for the transverse momentum distribution of protons (blue) and hyperons ($\Lambda$ in black, $\Sigma$ in green, $\Xi$ in red) in min.bias p+Au reactions at $E_{\rm lab}=$ 5-30 GeV with $|y_{\rm lab}|<0.5$.}
    \label{fig:dndpt_hyperons}
\end{figure*}

In Fig. \ref{fig:dndpt_hyperons} we show the transverse momentum distribution of protons and hyperons in the same p+Au reactions in the range $|y_{\rm lab}|<$0.5. The choice of the rapidity window is motivated by the fact that here most of the target nucleons are located and therefore the probability to create a nuclear hypercluster is largest, while still capturing the peak of the hyperon distribution. We observe that the proton transverse momentum distribution has a clear enhancement at small transverse momenta $0.0 < p_T < 0.5$ GeV/c which is from the fermi momentum of target. As expected the overlap in the transverse direction is also substantial and allows for copious production of hyper clusters.

Thus, we conclude that around $y_{\rm lab}=0-0.5$ we expect to find optimal conditions for the production of hypernuclei, due to the high baryon and hyperon yields and the still moderate longitudinal momenta of the hyperons which will increase the coalescence probability.  

\begin{figure*}[t!] 
    \centering
        \includegraphics[width=1.0\textwidth]{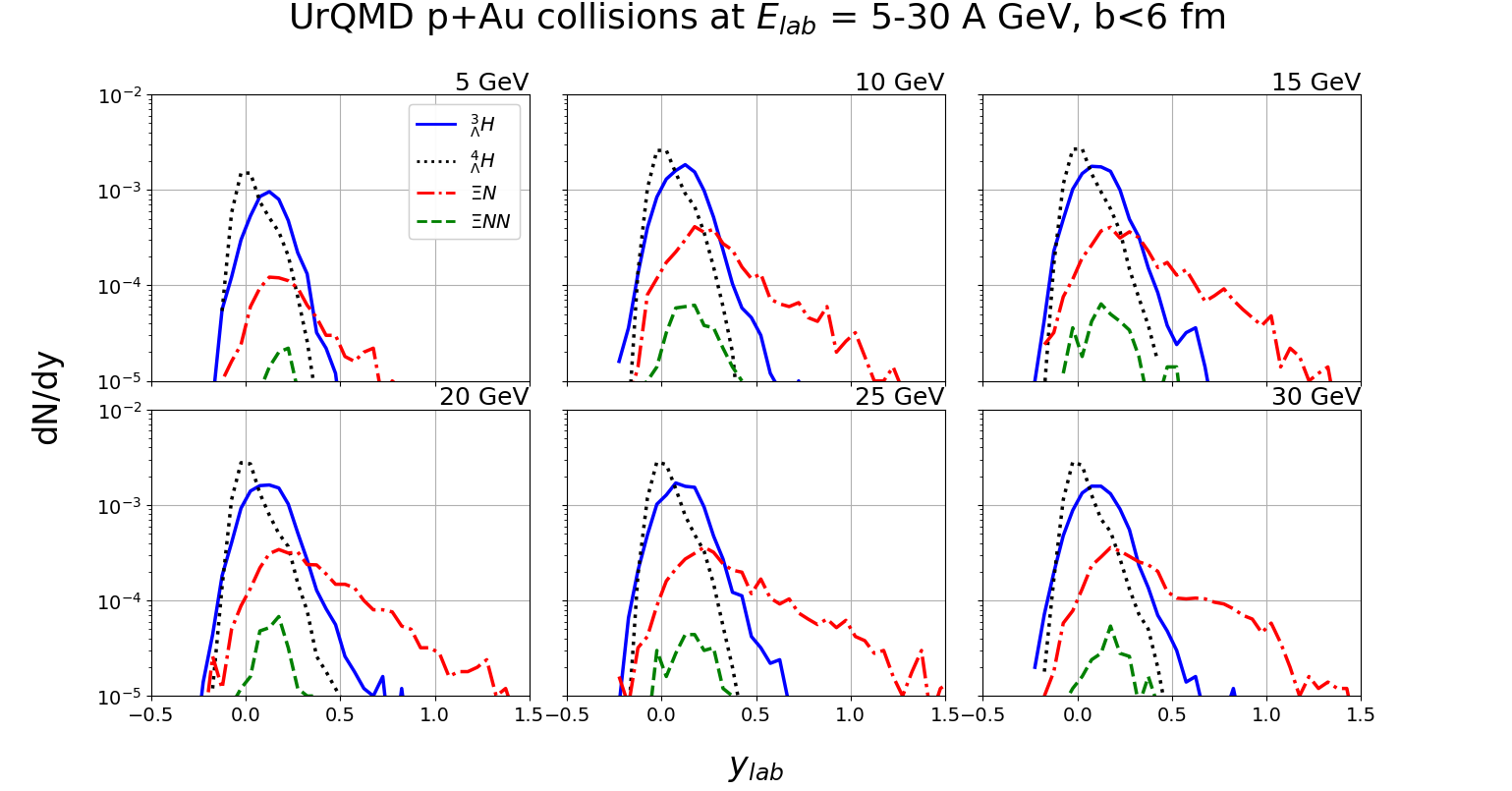}
    \caption{UrQMD predictions for the rapidity distribution of the hypernuclei $^{3}_{\Lambda}H$ (blue), $^{4}_{\Lambda}H$ (black), $\Xi$N (red) and $\Xi$NN (green) in min.bias p+Au reactions at $E_{\rm lab}=$ 5-30 GeV. }
    \label{fig:dndy_hyperclusters}
\end{figure*}


\begin{figure*}[t!] 
    \centering
        \includegraphics[width=1.0\textwidth]{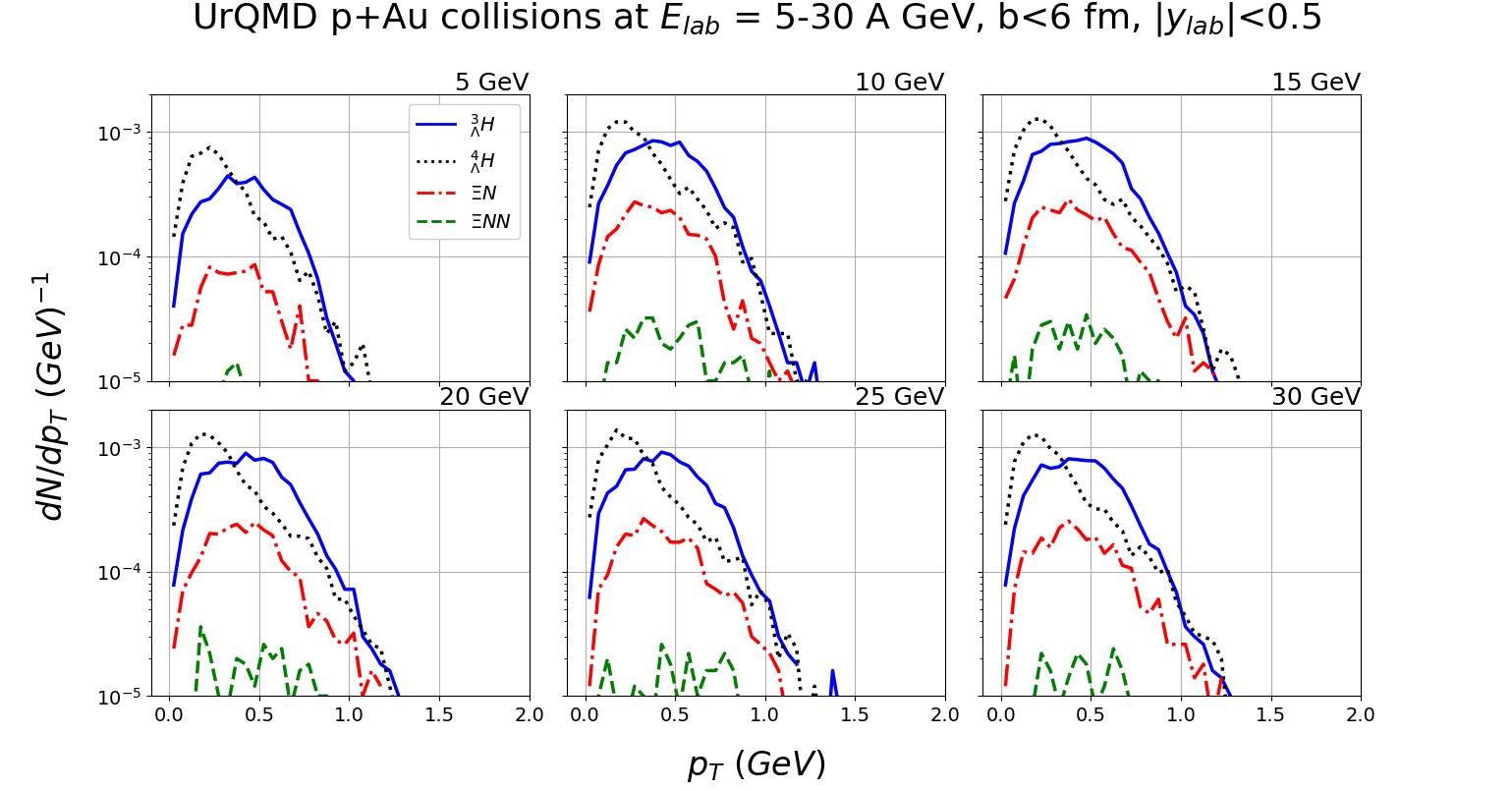}
    \caption{UrQMD predictions for the transverse momentum distribution of the hypernuclei $^{3}_{\Lambda}H$ (blue), $^{4}_{\Lambda}H$ (black), $\Xi$N (red) and $\Xi$NN (green) in min.bias p+Au reactions at $E_{lab}=$ 5-30 GeV within $|y_{\rm lab}|<0.5$.}
    \label{fig:dndpt_hyperclusters}
\end{figure*}

\section{Predictions for hypernuclei}
Let us finally turn to the production of hypernuclei. As discussed above, we employ the standard coalescence procedure implemented in UrQMD 4.0.  In Fig. \ref{fig:dndy_hyperclusters} we show the UrQMD predictions for the rapidity distribution of hypernuclei ($^{3}_{\Lambda}H$, $^{4}_{\Lambda}H$, $\Xi$N and $\Xi$NN) at $E_{\rm lab}=$ 5-30 GeV in min.bias p+Au reactions. As expected from the discussion above, we observe that the hypernucleus production shows a maximum in the target rapidity region. This becomes more pronounced if the hypernuclei consist of a higher number of non-strange baryons, while for the $\Xi$ hypernuclei also the forward rapidity region is still populated. A similar behavior was also observed in \cite{Chimruang:2026jqe} for charmed nuclei.

In Fig. \ref{fig:dndpt_hyperclusters} we show UrQMD predictions (min.bias p+Au reactions) for the transverse momentum distribution of hypernuclei at $y_{\rm lab}<$0.5 for the states $^{3}_{\Lambda}H$, $^{4}_{\Lambda}H$, $\Xi$N and $\Xi$NN. Here the maximum is found around a $p_T\approx 500$MeV, which should be accessible by the experiment, even if a low $p_T$ cut is applied to remove target remnants.

\section{Conclusion}
In this work we used the UrQMD model to calculate min.bias p+Au collisions in the FAIR energy range of $E_{\rm lab}=$ 5-30 GeV. We explored the correlated production of multiple hyperons in a single event to study the reach of CBM@FAIR and other planned experiments in terms of multi-hyperon hypernuclei. We found that up to 6 single strange hyperons might be present (with accessible statistics) in a single event, opening the route to produce and investigate multi-strange hypernuclei. Even multiple $\Xi$ production is possible and might allow to explore for the first time multi-$\Xi$ states. We have further explored the overlap of the produced hyperons and the nucleons (from the target nucleus) in momentum space and found that the target region provides an excellent environment to produce hypernuclei. This result is similar to previous studies using pion beams, although proton induced reactions have the advantage of a higher beam energy and higher luminosity which allows for the production of multiple hyperons or multi-strange hyperons. Finally, we have employed a coalescence approach to predict the yields and distributions of a broad variety of hypernuclei ranging from $^{3}_{\Lambda}H$, $^{4}_{\Lambda}H$, $\Xi$N to $\Xi$NN states. We predict that all these states are in reach of the CBM-experiment with multiplicities above $10^{-5}$ for the most exotic states and $10^{-3}$ for known states like the hyper-triton.

\section*{Acknowledgments}
This research has received funding support from the NSRF via the Program Management Unit for Human Resources \& Institutional Development, Research and Innovation [grant number B39G680010]. N.J. acknowledges support from the Development and Promotion of Science and Technology Talents Project (DPST), a Thai government scholarship.
The computational resources for this project were provided by the Center for Scientific Computing of the GU Frankfurt and the Goethe--HLR.

\end{document}